\documentclass[aps,prb,superscriptaddress,twocolumn]{revtex4}
\usepackage{graphicx}
\usepackage{times}
\usepackage{natbib}
\usepackage{color}
\usepackage{amsmath}
\usepackage[colorlinks,bookmarks=false,citecolor=blue,linkcolor=blue,urlcolor=blue]{hyperref}
\usepackage{threeparttable}
\long\def\symbolfootnote[#1]#2{\begingroup%
\def\thefootnote{\fnsymbol{footnote}}\footnote[#1]{#2}\endgroup}

\begin{document}


\title{Phase diagram with enhanced spin-glass region of mixed Ising / $XY$ magnets LiHo$_{x}$Er$_{1-x}$F$_{4}$}


\author{J. O. Piatek}
\email[]{julian.piatek@epfl.ch}
\author{B. Dalla Piazza}

\author{N. Nikseresht}
\author{N. Tsyrulin}
\affiliation{LQM, ICMP, Ecole Polytechnique F\'{e}d\'{e}rale de Lausanne, CH-1015 Lausanne, Switzerland}

\author{I. \v{Z}ivkovi\'{c}}
\affiliation{Institute of Physics, Bije\v{c}nika c. 46, HR-10000, Zagreb, Croatia}

\author{K. W. Kr\"{a}mer}
\affiliation{Department of Chemistry and Biochemistry, University of Bern, CH-3012 Bern, Switzerland}

\author{M. Laver}
\affiliation{Laboratory for Neutron Scattering, Paul Scherrer Institut, CH-5232 Villigen, Switzerland}
\affiliation{Department of Physics, Technical University of Denmark, DK-2800 Kgs. Lyngby, Denmark}

\author{K. Prokes}
\author{S. Mat’a\v{s}}
\affiliation{Helmholtz-Zentrum Berlin f\"{u}r Materialien und Energie,
Hahn-Meitner Platz 1, D-14109, Berlin, Germany}
\author{N. B. Christensen}
\affiliation{Department of Physics, Technical University of Denmark, DK-2800 Kgs. Lyngby, Denmark}

\author{H. M. R\o{}nnow}
\affiliation{LQM, ICMP, Ecole Polytechnique F\'{e}d\'{e}rale de Lausanne, CH-1015 Lausanne, Switzerland}

\date{\today}
\begin{abstract}
We present the experimental phase diagram of LiHo$_x$Er$_{1-x}$F$_4$, a dilution series of dipolar-coupled model magnets.The phase diagram was determined using a combination of AC susceptibility and neutron scattering. Three unique phases in addition to the Ising ferromagnet LiHoF$_4$ and the $XY$ anti-ferromagnet LiErF$_4$ have been identified. Below $x=0.86$ an embedded spin-glass phase is observed, where a spin-glass exists within the ferromagnetic structure. Below $x=0.57$ an Ising spin-glass is observed consisting of frozen needle-like clusters. For $x \sim 0.3 - 0.1$ an antiferromagnetically coupled spin-glass occurs. A reduction of $T_C(x)$ for the ferromagnet is observed which disobeys the mean-field predictions that worked for LiHo$_x$Y$_{1-x}$F$_4$.
\end{abstract}


\maketitle


\section{Introduction\label{introduction}}
When combinations of disorder and frustration are present in a magnetic system, the so called spin-glass state can arise. In such a state, the long range order is suppressed, but spins still exhibit spatial and temporal correlations. Over the last 30 years, much theoretical and experimental work has focused on spin-glasses, however materials with a well defined Hamiltonian are to this day something of a rarity. One candidate for a well defined spin-glass system has been LiHoF$_4$ diluted with non-magnetic yttrium~\cite{Reich1986,Reich1990,Wu1993,Tam2009}. The attractiveness of this system stems from the well described Hamiltonian of the Ising ferromagnet parent compound LiHoF$_4$~\cite{Hansen1975,Bitko1996,Ronnow2005,Ronnow2007}. Many of the interesting phenomena observed in LiHo$_x$Y$_{1-x}$F$_4$~\cite{Brooke1999,Brooke2001,Ghosh2003,Ronnow2005,Silevitch2007} are a direct consequence of the frustrated long range dipolar interaction between Ho$^{3+}$ moments. The combination of this frustrated interaction and the quenched disorder induced by the random Ho$^{3+}$ population on the rare earth site, makes this system a perfect candidate for spin-glass formation. From a theoretical point of view, the Ising anisotropy of the moments decreases the complexity, effectively to a simple anisotropic $s=1/2$ model, allowing for a large number of theoretical predictions to be made~\cite{tabei_induced_2006,schechter_quantum_2006,gingras_collective_2011}.

A recent theme in this system is the role of the off-diagonal dipolar terms in the Hamiltonian, whose effect is tuned by a magnetic field transverse to the Ising axis~\cite{schechter_quantum_2006,tabei_induced_2006}. It was discovered that in the presence of a small transverse field, LiHo$_x$Y$_{1-x}$F$_4$ is a perfect realization of the classical random field Ising magnet (RFIM)~\cite{tabei_induced_2006,schechter_liho_xy_1xf_4_2008}. One motivation for these studies was experimental non-linear AC susceptibility data in the presence of a transverse magnetic field, which suggested a T=0 first order spin-glass quantum phase transition~\cite{Wu1993,Ancona-Torres2008}. A second group found no evidence of either a classical or quantum spin-glass state~\cite{Jonsson2007}. A third group found very strong evidence of a classical spin-glass~\cite{quilliam_2007,quilliam_2008,quilliam_2012}, but did not investigate its behavior in the presence a transverse field. On the theoretical side, two independent groups reach the conclusion that the presence a transverse field induces random-fields which destroy true spin-glass order~\cite{tabei_induced_2006,schechter_quantum_2006,schechter_liho_xy_1xf_4_2008}. Furthermore it has been demonstrated~\cite{tabei_induced_2006} that the observed non-linear susceptibility~\cite{Wu1993} is consistent with this spin-glass-like state.

\begin{figure}[tbp]
\includegraphics[]{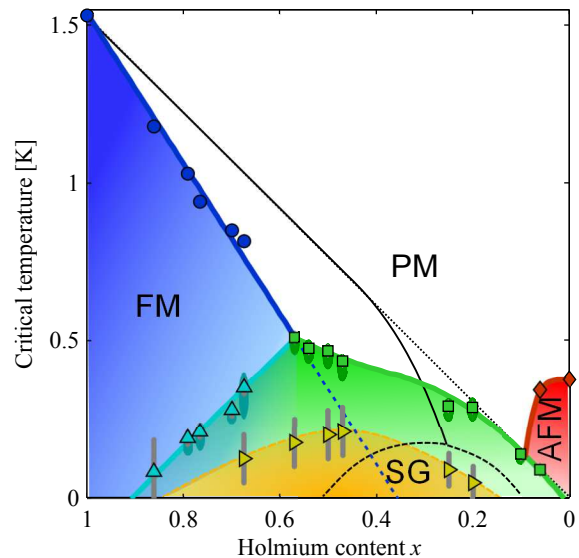}
\caption{(Color online). Experimental phase diagram of LiHo$_x$Er$_{1-x}$F$_4$ with the phase diagram of LiHo$_x$Y$_{1-x}$F$_4$ (black line) overlaid to highlight the extended glassy region observed here. Blue circles and red diamonds show respectively ferromagnetic ($T_C$) and antiferromagnetic ($T_N$) order. Green squares indicate a spin-glass freezing ($T_{\text{f}}$) from a paramagnetic state. Secondary glass-like transitions from either a ferromagnetic or spin-glass state are shows as light blue triangles and yellow triangles respectively. The ellipses show the range of $T_{\text{f}}$ in the frequency range measured for each sample (typically $1$ Hz to $4$ kHz). Lines are guides to the eye.}
\label{phasediag}
\end{figure}

The key idea of this paper is that introducing Er into the parent compound in place of Y should increase frustration due to the $XY$ anisotropy observed in LiErF$_4$~\cite{Beauvillain1977,Magarino1980,kraemer-lierf4-2012}. The addition of these off-diagonal terms in the Hamiltonian could drive the onset of glassy behavior up to higher temperatures and extend the range of concentrations where they are visible. Moreover the resulting compounds are in the class of mixed anisotropy systems, where rich phase diagrams including co-existing magnetic phases have been both theoretically predicted~\cite{fishman_phase_1978,fishman_phase_1979,fishman_phase_1980} and experimentally observed~\cite{Wong_1980,Wong_1983}. This paper presents AC susceptibility and neutron diffraction studies on LiHo$_x$Er$_{1-x}$F$_4$ for 15 concentrations from $x=1$ to $x=0$. We establish a phase diagram with several striking features compared both to LiHo$_x$Y$_{1-x}$F$_4$ and to simple mean-field (MF) expectations for LiHo$_x$Er$_{1-x}$F$_4$: a) The spin-glass region is greatly extended, both in terms of $x$ and in temperature, up to a maximum $T_{\text{f}} > 0.5~K$; b) $T_C$ for ferromagnetic order decreases faster than for Y-dilution and in contrast to MF predictions; c) the antiferromagnetic order found in pure LiErF$_4$ is destroyed by just 10\% of Ho. 

The experimentally determined phase diagram for  LiHo$_x$Er$_{1-x}$F$_4$ is shown in Fig.~\ref{phasediag}. Blue circles indicate the ferromagnetic $T_C$, green squares the spin-glass $T_{\text{f}}$ at $1$ kHz and red diamonds the antiferromagnetic $T_N$. Additional freezing transitions which occur within a ferromagnetic or spin-glass state are indicated by the light blue and yellow triangles respectively. The light blue and green ellipses show the range of $T_{\text{f}}$ observed in the frequency range measured for each sample. Moving from large to small $x$, three phases in addition to the  Ising ferromagnet for LiHoF$_4$ and $XY$ antiferromagnet in the case of LiErF$_4$ are observed. Already at $x=0.86$ at very low temperatures there are indications of some form of spin-glass behavior. The temperature of the onset of the glassy state increases steadily with decreasing $x$. Once $x$ falls below a critical value, of around $x=0.57$, the long range ferromagnetic order is completely suppressed and replaced by a spin-glass phase. The maximum observed $T_{\text{f}}$ is found for $x=0.57$ and gradually decreases until antiferromagnetic order occurs for LiErF$_4$. The phase diagram for LiHo$_x$Y$_{1-x}$F$_4$ is shown in black and highlights the extended glassy region observed here.

One challenge in experimental studies of LiHo$_x$Y$_{1-x}$F$_4$ is the low temperatures of the spin-glass phase at $T \lesssim 150$~mK. This is compounded by the poor thermal conductivity and large specific heat of the samples, which make it difficult at best to be certain of thermal equilibrium. From the details expounded below, it can be seen that extraordinary steps were taken to ensure the thermal equilibrium of our LiHo$_x$Er$_{1-x}$F$_4$ samples.

The layout of this paper is as follows. Section~\ref{susceptibility} outlines the AC susceptibility experiment and describes the results. Section~\ref{neutrons} discusses neutron diffraction results obtained on samples with $x=0.79, 0.5$ and $0.25$. Both sets of results are interpreted in the framework of mean-field calculations in Section~\ref{Discussion}. The final section is dedicated to discussion of the results in a more general context and conclusions.

\section{\label{susceptibility}AC Susceptibility}

LiHo$_x$Er$_{1-x}$F$_4$ samples were prepared from mixtures of LiF and $M$F$_3$ in a 53:47 (Li:$M$) molar ratio with $x$ the ratio of HoF$_3$ to ErF$_3$. The mixture of salts was melted in a glassy carbon Bridgman ampule at a temperature of 880$^\circ$ C in an inert gas atmosphere. The melt was subsequently slowly cooled over 7 days. The purity of the resulting polycrystals was checked by powder X-ray diffraction and the Ho to Er ratio verified by EDX measurements with an accuracy of $\pm 0.5\%$.

Small polycrystals of LiHo$_x$Er$_{1-x}$F$_4$ were ground to powder and then mixed with Stycast W19 to ensure good thermal contact. The resulting paste was pressed into a 20 mm long 2 mm diameter mold and four 200$\mu m$ diameter copper wires were inserted into the mixture which was then baked to cure the Stycast. The copper wires were attached directly to the thermometer housing on the mixing chamber, to ensure the best possible thermalization. Complex AC susceptibility measurements were carried out in an Oxford Kelvinox 25 dilution refrigerator, using coaxially compensated mutual inductance coils. The primary coils are supplied with an AC current using a Keithley 6221 AC current source and the signal induced in the secondary coils is measured using a Signal Recovery 7265 lock-in amplifier.  The measurements used an AC excitation current of $10\mu A$, corresponding to a field of 42~mOe, in the range of 1 Hz -  4 kHz.

Before looking at the results in detail, some concepts related to spin-glasses must be introduced. The spin-glass state is typically characterized by a sharp cusp in the real signal of the AC susceptibility, around the freezing temperature $T_{\text{f}}$, which is rounded out by the smallest fields~\cite{mydosh1993}. In the case of insulating spin-glasses the imaginary susceptibility is dominated by the spin-glass dynamics. $\chi^{\prime\prime}$ begins to increase at $T>T_{\text{f}}$, has an inflection point which coincides with the peak in $\chi^\prime$ and peaks shortly afterwards before dropping down to zero at low temperatures. The temperature at which the $\chi^{\prime}$ cusp occurs depends on the frequency of the AC driving field, as correlated spins in different local environments with different relaxation times will freeze out at different temperatures. The frequency dependence of $T_{\text{f}}$ is logarithmic for almost all experimentally accessible frequencies ($10^{-3} - 10^{10}$ Hz) and can be well described by an Arrhenius law:

\begin{equation}
f=1/\tau_{\text{char}}=f_0\exp(-E_a/k_BT),
\label{eq:Arrhenius}
\end{equation}

where $E_a$ and $f_0$ are the energy barrier and characteristic frequency in the case of a superparamagnet. One way of separating a superparamagnet, where spins form non-interacting clusters, from a spin-glass, which contains long range correlations is by looking at the frequency sensitivity of $T_{\text{f}}(f)$~\cite{Mahendiran2003,Singh2008}:

\begin{equation}
\mathcal{K}=\frac{\Delta T_{\text{f}}}{T_{\text{f}}\Delta\log_{10}(f)}.
\label{eq:freq_sens}
\end{equation}

Generally metallic canonical spin-glasses have values of $\mathcal{K} \leq 0.01$, insulating spin-glasses have values in the region $0.01 \leq \mathcal{K} \leq 0.1$ and shifts higher than this are typical of superparamagnets. 

An alternate explanation of the frequency dependence of $T_{\text{f}}$ is that of a critical scaling law associated with a spin-glass phase transition at finite temperature $T_g$. This explanation was introduced when a divergence from Arrhenius law was observed at $T_{\text{f}} \rightarrow T_g$ in some spin-glass systems~\cite{paulsen_1987}. The critical scaling is usually expressed as~\cite{Fischer1991}:

\begin{equation}
\tau=\tau_0(T/T_g-1)^{-z\nu},
\label{eq:scaling}
\end{equation}

where $\tau$ is the characteristic spin relaxation time at temperature $T$, $T_g$ is the glass temperature and the product $z\nu$ is the dynamical exponent. This form of dynamical scaling is supported by calculation on Ising spin-glass models, where Monte Carlo simulations~\cite{Ogielski1985,Ogielski1985_1} find critical scaling with $z\nu\approx7$. Care should be taken when referring to the different freezing temperatures $T_{\text{f}}$ and $T_g$; $T_{\text{f}}$ is the frequency dependent temperature at which the spin-glass transition is observed, whereas $T_g$ is the zero-frequency spin-glass transition temperature. As will be demonstrated, it can be difficult to distinguish between these two dynamical situations in the range of experimentally accessible frequencies. A final complication in LiHo$_x$Er$_{1-x}$F$_4$ is the large anisotropy of the Ising moments which can prevent the system reaching equilibrium, resulting in a thermally activated spin-glass\cite{barbara_activated_2007}.

\begin{figure}[p]
\includegraphics[width=8cm]{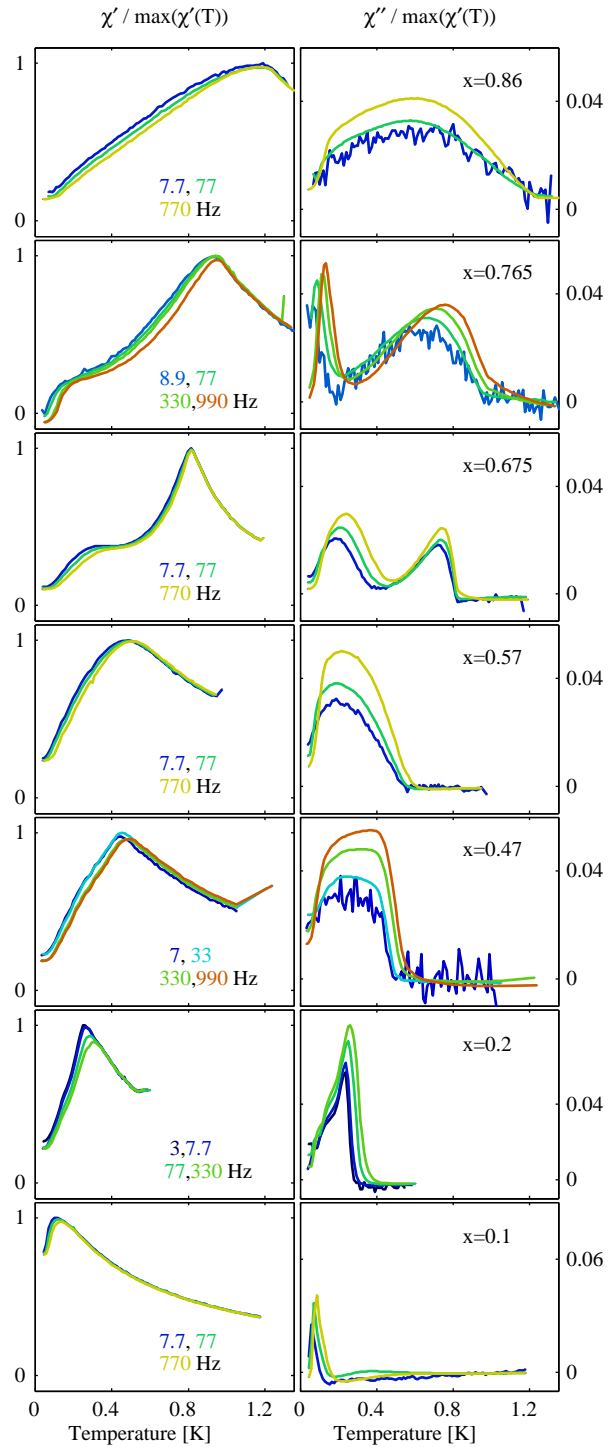}
\caption{(Color online). AC susceptibility data for LiHo$_x$Er$_{1-x}$F$_4$. Moving from high to low $x$, the ferromagnetic peak is suppressed by the additional Er content, disappearing completely at $x=0.57$. Frequency dependent behavior is seen in all samples, in the form of an embedded spin-glass for $0.68< x \lesssim 0.9$ and more typical spin-glasses for $x < 0.57$. }
\label{overview}
\end{figure}

An overview of the real, $\chi^{\prime} (T)$, and the imaginary, $\chi^{\prime\prime}(T)$, components of the complex AC susceptibility is presented for selected compositions in Fig.~\ref{overview}. Starting from large $x$, a frequency independent peak in $\chi^{\prime}$ and concomitant onset of $\chi^{\prime\prime}$ is seen just below 1~K, which we will later show to be ferromagnetic order. At lower temperatures there is a frequency dependent drop in $\chi^{\prime}$ which appears to correspond to a peak in $\chi^{\prime\prime}$. As $x$ decreases T$_C$ decreases linearly with $x$ down to $x=0.57$, where long range order is no longer observed. The peak in $\chi^{\prime}$ broadens and the temperature of the peak now depends on frequency, indicating a slowing down process consistent with a spin-glass state. This frequency dependent peak continues all the way down to $x=0.1$, with the features in the susceptibility constantly evolving with $x$.

\subsection{Ferromagnetic region}

\begin{figure}[tbp]
\includegraphics[width=8cm]{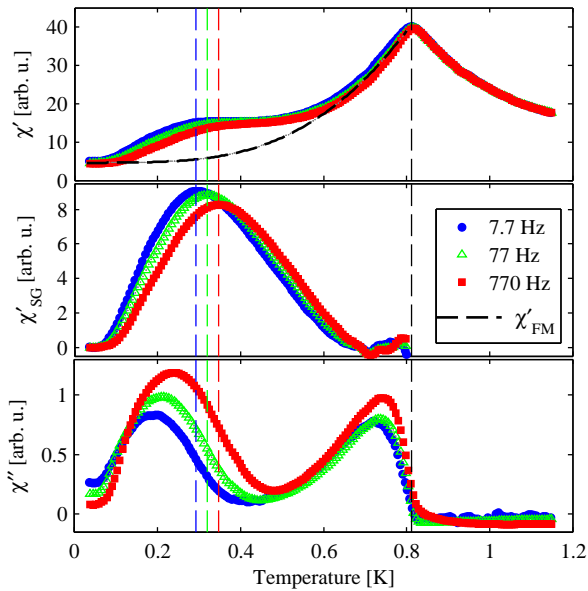}
\caption{(Color online). Temperature scans of LiHo$_{0.675}$Er$_{0.325}$F$_4$ taken at 7.7, 77 and 770 Hz. The peak in $\chi'$ at 0.815~K is $T_C$ of the ferromagnetic long range order, and the frequency dependence seen at ~$T=0.2$K$ - 0.4$K corresponds to a spin glass transition. $T_{\text{f}}$ is determined by subtracting a ferromagnetic background signal (dashed line) from the measured curve, leaving the usual peak in the susceptibility (middle panel).}
\label{FM+SG}
\end{figure}

We illustrate the analysis of the ferromagnetic region in detail for $x=0.675$ in Fig.~\ref{FM+SG}, which shows susceptibility curves taken at $7.7$, $77$ and $770$ Hz. At $T=0.815$~K there is simultaneously a peak in $\chi^{\prime}$ (top panel) and a kink of $\chi^{\prime\prime}$ (bottom panel). This is a clear indication of a ferromagnetic transition where $\chi^{\prime}$ diverges and domain wall motion causes a sudden increase in $\chi^{\prime\prime}$. This high temperature behavior is essentially the same behavior as observed for LiHoF$_4$~\cite{Bitko1996}, with the exception that $T_C$ has now decreased to lower temperatures. At lower temperatures, there is a clear frequency dependent behavior in both $\chi^{\prime}(T)$ and $\chi^{\prime\prime}(T)$. The frequency dependent peak observed in $\chi^{\prime\prime}$ just below $T_C$ is probably due to domain wall motion in the ferromagnet. The lower temperature peak in $\chi^{\prime\prime}$ at temperatures around 250 mK is likely due to the freezing out of moments not in the ferromagnetic state.

To better determine the behavior of the frequency dependent features we note that the presence and dynamics of domain walls in the sample, inferred from the peak  $\chi^{\prime\prime}$ just below T$_C$, will give rise to a signal in $\chi^{\prime}$. The observed $\chi^{\prime}$ is therefore considered to be the superposition of two physically distinct phenomena, a high temperature signal from the ferromagnet and a low temperature signal from a spin-glass. The ferromagnetic component of the susceptibility is estimated by fitting the data with an exponential function in the temperature range where the signal from the freezing is assumed to be negligible (in this case $0.03$ K $<T<0.075$ K and $0.6$ K $<T<0.86$ K). Subtracting the fitted component from the measured susceptibility reveals a signal believed to be coming from the spin-glass (middle panel), which peaks at roughly the same position as the inflection point of $\chi^{\prime\prime}$, as is expected for a typical spin-glass.

\begin{figure}[tbp]
\includegraphics[width=8cm]{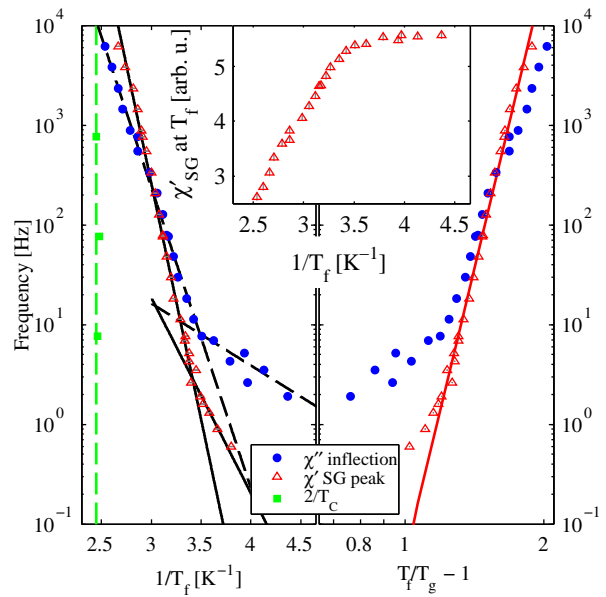}
\caption{(Color online). [left] Frequency dependence of $T_{\text{f}}$ for $x=0.675$ with a fit to the Arrhenius law. Comparison of the frequency dependence of $T_{\text{f}}$ if determined by the inflection point before the peak in $\chi^{\prime\prime}$ or the peak in $\chi^{\prime}$ after subtraction of the background signal from the ferromagnet. To demonstrate the frequency independence of $T_C=0.815$~K, $2/T_C$ is also plotted. [right] Scaling law for the same data sets assuming a zero frequency glass transition temperature $T_g$. The inset shows the temperature dependence of the $\chi^{\prime}_{SG}$ peak amplitude.}
\label{FM+SG_Fdep}
\end{figure}

The frequency dependence of $T_{\text{f}}$ is traditionally compared to either an Arrhenius law or a scaling law~\cite{mydosh1993}. Avoiding prejudice between these two paradigms for spin-glasses, we show in the left panel of Fig.~\ref{FM+SG_Fdep}  an Arrhenius type plot and in the right panel a scaling law plot. In order to carry out the scaling law analysis in a reproducible fashion, the value of $T_g$ must be determined and fixed. As there is no clear indication of a low frequency saturation in $T_{\text{f}}$, $T_g$ is determined by fitting eq.~\ref{eq:scaling} for the entire possible range of $T_g$ ($0<T_g<T{_\text{f}\text{(min)}}$). The trial temperature which gives the best fit to this equation is assumed to be the zero frequency $T_g$. The figure shows the dynamics of $T_{\text{f}}$ as extracted from the inflection point in $\chi^{\prime\prime}$ (blue circles) and from peak in $\chi^{\prime}$ (red triangles) revealed by subtracting the ferromagnetic component.  To illustrate the frequency independence of $T_C$, the green squares show $2/T_C$. 

Focusing first on the Arrhenius plot, a value of $\mathcal{K}=0.02$ is extracted for the high temperature (and high frequency) region. The extracted value of the attempt frequency, $f_0=4\times10^{10}$~Hz, gives an indication that the dynamics could be well described by an Arrhenius law. At lower temperatures and longer time-scales, the Arrhenius behavior is lost and it appears that the system crosses over into a different set of dynamics, which can be explained with a new Arrhenius law with $\mathcal{K}=0.04$. An examination of the peak in the real susceptibility (inset of Fig.\ref{FM+SG_Fdep}) also points towards an underlying change in behavior; the peak amplitude becomes temperature independent at the same temperature of the cross-over in slope. The dynamics can be equally well explained by critical scaling, with $T_g = 0.13 \pm 0.002$~K, $\tau_0 = 3.5 \pm 0.5$~s and $z\nu = 19 \pm 0.5$. The very large values of $\tau_0$ and $z\nu$ coupled with the low temperature deviation from the fit seem to imply that the system does not follow a dynamic scaling law.

\subsection{Large $x$ Spin Glass Region}

As $x$ is decreased further, the ferromagnetic order disappears completely, and the system shows very broad frequency dependent features. Given the relatively similar amounts of Ho and Er, and the behavior of the large $x$ phase, it seems plausible that there are in fact two transitions which constitute the broad signal. 

\begin{figure}[tbp]
\includegraphics[width=8cm]{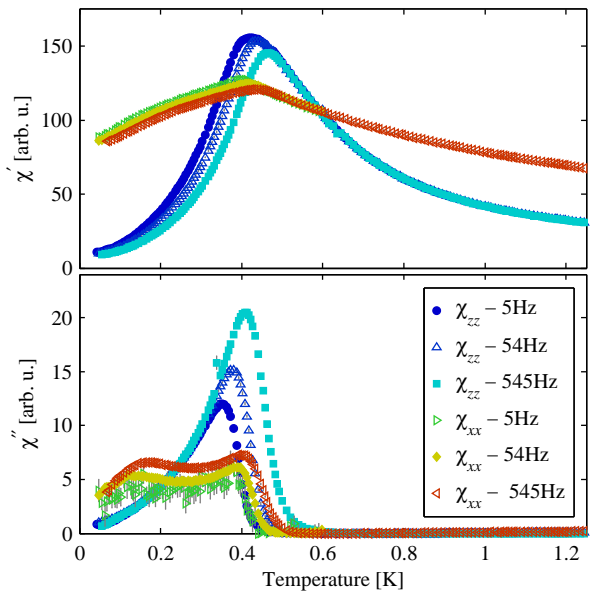}
\caption{(Color online). Temperature scans of single crystals of LiHo$_{0.50}$Er$_{0.50}$F$_{4}$ comparing $\chi_{zz}$ and $\chi_{xx}$. In order to better compare the two data sets $\chi_{xx}^{\prime}$ has been scaled by a factor of 20 and $\chi_{xx}^{\prime\prime}$  by a factor of 50.}
\label{50_zz_xx}
\end{figure}

In the case of $x=0.50$, the susceptibility has been measured on single crystals in addition to the powder samples. Given the very high anisotropy of both Ho and Er moments and their different nature, it is possible to partially separate their contributions, as $\chi_{zz}$ is predominantly sensitive to the Ising moments of Ho and $\chi_{xx}$ predominantly sensitive to the XY Er moments. A comparison of these two susceptibilities is shown in Fig.~\ref{50_zz_xx}. The susceptibilities measured along different crystallographic directions show strikingly different behavior. At T=$0.450\pm 0.02$~K, $\chi^{\prime}$ peaks in both measurements, with the peak position depending on the frequency of the measurement, indicating the freezing of the moments. Below this temperature, $\chi_{zz}$ decreases relatively quickly, as expected for an Ising spin-glass, whereas $\chi_{xx}$ remains relatively broad and does not decrease significantly. The imaginary component of the susceptibility shows similar differences, where two frequency dependent peaks are clearly visible in $\chi_{xx}^{\prime\prime}$.

\begin{figure}[tbp]
\includegraphics[width=8cm]{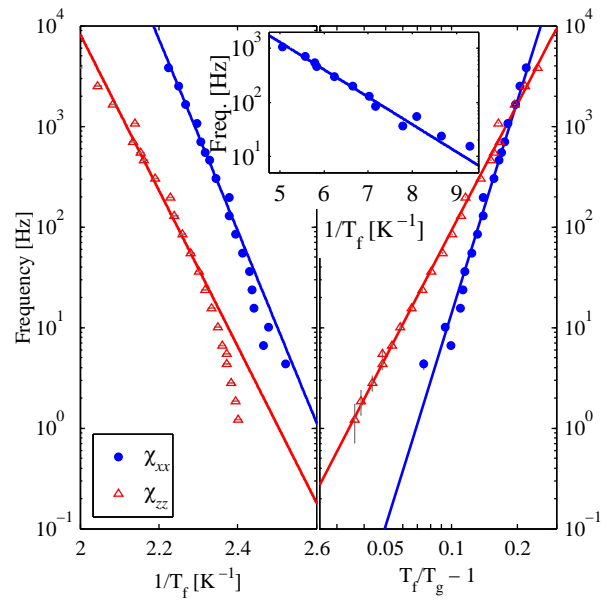}
\caption{(Color online). Frequency dependence of $T_{\text{f}}$ in LiHo$_{0.50}$Er$_{0.50}$F$_{4}$
expressed in terms of [left] Arrhenius behavior and [right] a dynamic scaling law. The inset shows the frequency dependence of the low temperature peak in $\chi_{xx}^{\prime\prime}$ in the range of frequencies where the peak was extractable.}
\label{50_freq}
\end{figure}

As was the case for $x=0.675$, the frequency dependence of $T_{\text{f}}$ is fitted both by Arrhenius activated dynamics and critical scaling, as shown in Fig.~{\ref{50_freq}}. There is a low temperature deviation from Arrhenius dynamics in the opposite direction to the $x=0.675$ sample; in this case the freezing temperature tends towards a zero-frequency value. This is reflected in the critical scaling dynamics, where there is no visible deviation from the expected fit. The large value of $f_0\approx10^{20}$~Hz would be a further argument that the system undergoes a true spin-glass phase transition and is therefore described by critical scaling dynamics. The lowest temperature peak in $\chi_{xx}^{\prime\prime}$ is believed to be due to a second glass transition, whose freezing temperature is taken as the peak in susceptibility. As the peak is small compared to the background signal it is only possible to extract $T_\text{f}$ between 10 and 1000 Hz. In this frequency range the dynamics are well explained by an Arrhenius law, as can be seen in the inset of Fig.~\ref{50_freq}.

\begin{figure}[tbp]
\includegraphics[width=8cm]{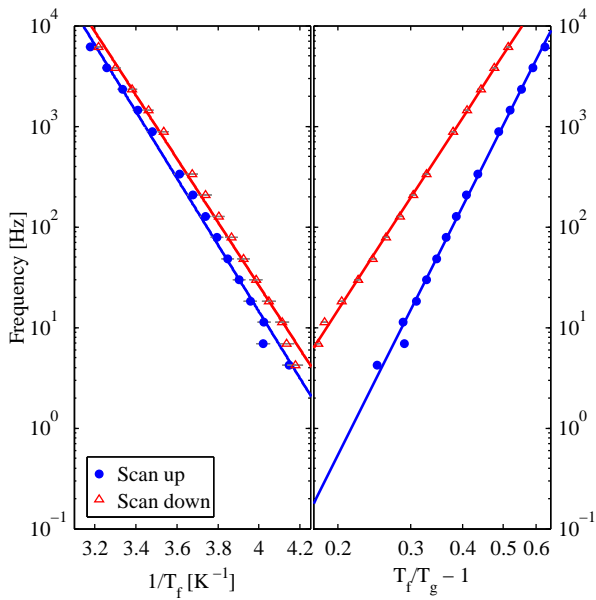}
\caption{(Color online). Frequency dependence of $T_{\text{f}}$ in LiHo$_{0.25}$Er$_{0.75}$F$_{4}$
expressed in terms of {[}left{]} Arrhenius behavior and {[}right{]}
critical scaling. The data compares $T_{\text{f}}$ extracted from
scans ramping both up and down in temperature, which are offset by
5 mK, probably due to thermalization issues.}
\label{SG_fdep}
\end{figure}

\subsection{Small $x$ Spin-Glass Region}

As $x$ decreases further, the spin-glass freezing temperature continues to slowly decrease and the features in the susceptibility sharpen. At $x=0.25$ the susceptibility has sharpened to the point that both $\chi^{\prime}$ and $\chi^{\prime\prime}$ resemble those found in typical spin-glasses~\cite{mydosh1993}. The frequency dependence of this sample is shown in Fig.~\ref{SG_fdep}, which plots the dependence of $T_{\text{f}}$ taken for temperature scans when ramping either low to high temperatures (blue circles) or from high temperatures to low ones (red triangles). The difference between the two scans give an idea of the maximum offset due to insufficient thermalization, which is found to be on the order of 5 mK. In this sample, both Arrhenius and critical scaling dynamics produce fits of equal quality, with no clear variation from either law immediately visible. A comparison of the relevant parameters from the Arrhenius and critical scaling fits does not shed any more insight into the nature of the dynamics in this compound – both are within the expected ranges for the respective fit type.

\begin{table}

\caption{Summary of transition temperatures and spin-glass parameters related to dynamics in LiHo$_{x}$Er$_{1-x}$F$_{4}$\label{tab_summary}. All temperatures are in Kelvin.}
\begin{centering}
{\small \begin{threeparttable}[b]
	\renewcommand{\arraystretch}{1.2}
	\begin{tabular}{c|c|ccc|ccc}
	\hline
	\hline
	$x$ & $T_{c}$ & $T_{f}$(1kHz) & $f_{0}$ [Hz] & $\mathcal{K}$ & $T_{g}$ & $z\nu$ & $\tau_{0}$ {[}s{]} \tabularnewline
	\hline
	1 & 1.53 & N/A & N/A & N/A & N/A & N/A & N/A \tabularnewline
	0.860 & 1.20 & 0.083 & $4\times10^{5}$ & 0.077 & - & - & - \tabularnewline
	0.790 & 1.04 & 0.191 & $4\times10^{8}$ & 0.044 & 0.043 & 12.9 & $1.3\times10^{3}$ \tabularnewline
	0.765 & 0.95 & 0.197 & $2\times10^{11}$ & 0.080 & 0.174 & 2.2 & $1.6\times10^{-6}$ \tabularnewline
	0.700\tnote{1} $^{}$& 0.85 & 0.280 & $2\times10^{15}$ & 0.025 & 0.156 & 10 & $1.3\times10^{-5}$ \tabularnewline
	0.675 & 0.815 & 0.347 & $4\times10^{10}$ & 0.027 & 0.130 & 19 & 3.5 \tabularnewline
	0.570 & N/A & 0.510 & $5\times10^{34}$ & 0.024 & 0.430 & 9.0 & $6.2\times10^{-11}$ \tabularnewline
	0.540 & N/A & 0.474 & $3\times10^{20}$ & 0.046 & 0.396 & 6.4 & $4.3\times10^{-9}$ \tabularnewline
	0.500\tnote{2} & N/A & 0.471 & $4\times10^{19}$ & 0.049 & 0.405 & 4.1 & $1.5\times10^{-7}$ \tabularnewline
	0.500\tnote{1} & N/A & 0.437 & $1\times10^{24}$ & 0.042 & 0.369 & 7.0 & $1.0\times10^{-9}$ \tabularnewline
	0.470 & N/A & 0.434 & $2\times10^{20}$ & 0.054 & 0.299 & 6.2 & $6.4\times10^{-9}$ \tabularnewline
	0.250 & N/A & 0.288 & $3\times10^{13}$  & 0.069 & 0.199 & 8.2 & $3.7\times10^{-7}$\tabularnewline
	0.200 & N/A & 0.286 & $3\times10^{13}$ & 0.070 & 0.196 & 7.5 & $4.2\times10^{-7}$\tabularnewline
	0.100 & N/A & 0.138 & $3\times10^{11}$ & 0.080 & 0.079 & 8.6 & $1.3\times10^{-5}$\tabularnewline
	0.060 & 0.342 & (0.15) & - & - & - & - & - \tabularnewline
	0 & 0.375 & N/A & N/A & N/A & N/A & N/A & N/A \tabularnewline
	\hline 
	\hline
\end{tabular}
\begin{tablenotes}
\item[1] Single crystal measurements of $\chi_{xx}$. 
\item[2] Single crystal measurements of $\chi_{zz}$. 
\end{tablenotes} 
\end{threeparttable}}
\par\end{centering}{\small \par}

\end{table}

All of the samples measured have been analyzed in a similar manner to these three examples, with the data being used to build up the phase diagram shown in Fig~\ref{phasediag}. The parameters relating to the phase transitions, $T_C$, $T_{\text{f}(1 kHz)}$, and $T_{\text{g}}$ and dynamics, $\mathcal{K}$, $z\nu$, and $\tau_0$, have been extracted and are summarized in Table~\ref{tab_summary}.

\section{\label{neutrons}Neutron scattering}

Neutron scattering measures the correlation function of the moments, therefore complementing the AC susceptibility measurements which measure the spatially averaged dynamics of the system. In particular, neutron scattering allows for the determination of spatial correlations and the directions of the magnetic moments involved in the spin-glass state. Furthermore, neutron scattering can be used to discern whether the correlations between moments are ferromagnetic or antiferromagnetic in nature.

\begin{figure}[tbp]
\includegraphics[width=8cm]{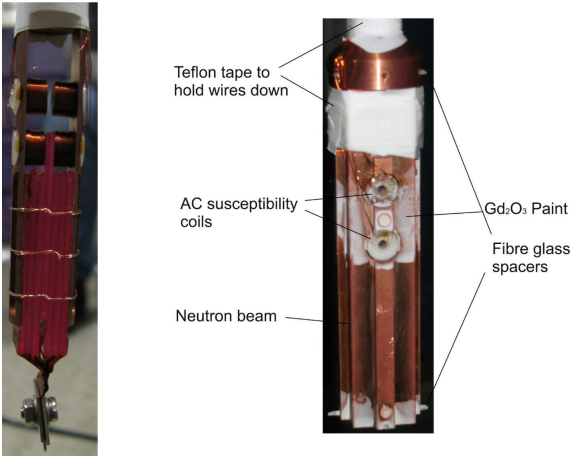}
\caption{(Color online). Photo of the sample holder used for the measurements on LiHo$_{0.25}$Er$_{0.75}$F$_4$. (left) Preliminary sample holder showing the configuration of the individual blades in the sample holder. (right) Sample holder used for experiments made from a single piece of OFHC copper. Each blade has been sputtered with gold and thermalized to the sample holder via a 12 mm x 25 $\mu$m copper foil. The central blade was cut 10 mm longer than the others and sits inside a split coils susceptometer to allow for in-situ measurements of AC susceptibility.}
\label{s_holder}
\end{figure}

Neutron scattering measurements have been carried out on three samples, with $x=0.79,0.5$~and~$0.25$. Due to the poor thermal conductivity of the Li$M$F$_4$ compounds, particular care has been taken in ensuring the best possible thermal conductivity. Large single crystals were cut into thin slices roughly 1.5 mm thick and individually gold sputtered creating a layer 2-3 $\mu$m thick. These blades were then placed inside a solid walled box made from a single piece of oxygen free high conductivity (OFHC) copper, with a thin copper foil strip placed between each blade. The copper foil strips are attached to the sample holder and a lid is screwed on to the box, pressing the blades together and thus ensuring a good thermal contact. In the case of $x=0.25$ AC susceptibility was measured $in-situ$ during the neutron scattering experiment using a homemade split coil AC susceptometer. Fig.~\ref{s_holder} shows a picture of the mounted sample holder for this experiment (right) and a preliminary sample holder (left), where the configuration of the blades of crystal is clearly visible. For the measured sample, the central blade was cut 10 mm longer than the rest, and the split coil susceptometer was placed on either side of this protrusion to measure the AC susceptibility.

\subsection{Ferromagnetic region}

Measurements on $x=0.79$ were carried out on the E4 thermal neutron diffractometer at HZB in Germany on five blades cut from a single crystal with total dimensions of $5 \times 5 \times 30$ mm. The sample was placed inside a dilution fridge and vertical-axis superconducting magnet, allowing for temperatures down to 50 mK and fields up to 5 T. The field direction was perpendicular to the c-axis along the $(h \bar {h}0)$ direction in order to study the possible transverse field Ising model (TFIM) quantum phase transition. The scattering plane was the $(hh0)-(00l)$ plane and the incoming beam of neutrons was monochromated to a wavevector of $k_i = 2.59~\text{\AA} ^{-1}$.

\begin{figure}[tbp]
\includegraphics[width=8cm]{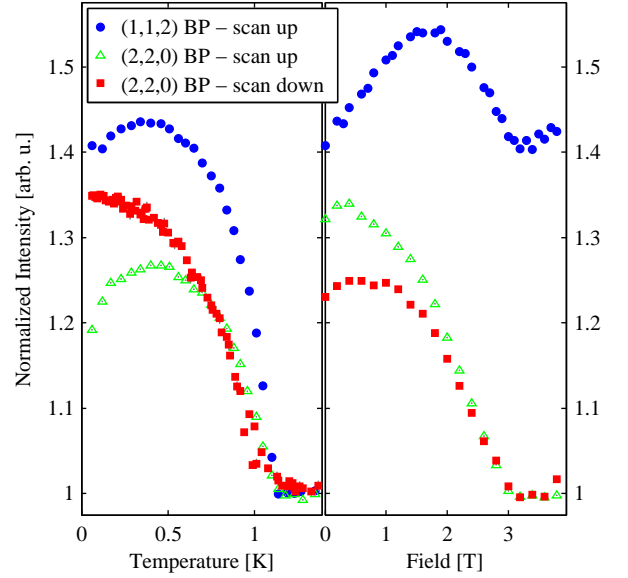}
\caption{(Color online). [left] Temperature and [right] field dependence of the (2,2,0) and (1,1,2) Bragg peaks in LiHo$_{0.79}$ Er$_{0.21}$ F$_{4}$, indicating the presence of long range ferromagnetic order. The data has been normalized to the nuclear Bragg peak intensity and the magnetic scattering from the (1,1,2) Bragg peak been scaled down by a factor of 5 relative the the (2,2,0) Bragg peak.}
\label{80_e4}
\end{figure}

The long range ferromagnetic order has been confirmed by measuring the temperature dependence of the $(2,2,0)$ and $(1,1,2)$ nuclear Bragg peaks, shown in the left panel of Fig.~\ref{80_e4}. The scan up is measured after having been at high field, so can therefore be thought of as a high field-cooled (HFC) state, whereas the scan down is a zero field-cooled (ZFC) state. As $T$ drops below $T_C$ the intensity of the Bragg peak begins to increase, indicating the build up of ferromagnetic order. The Bragg peak could be fitted using a resolution limited Gaussian lineshape, which indicates that the ferromagnetic correlations are long ranged. There is also a hysteresis between the HFC and ZFC states and the ordered moment in the ferromagnet is permanently reduced by being at large transverse fields.

The field dependence of the ferromagnetic signal has also been measured for these two Bragg peaks and is shown in the right panel of Fig.~\ref{80_e4}. For the (2,2,0) Bragg peak the picture is relatively simple, the application of the transverse field gradually destroys the long-range ferromagnetic order. The (1,1,2) peak intensity first increases, then decreases and finally increases above $H_C$. This behavior is likely a combination of the spins being polarized by the field (increasing intensity) and destruction of the c-axis correlations (decreasing intensity). Such behavior is not seen in the (2,2,0) Bragg peak as the moments are polarized parallel to the scattering vector and as such have a vanishing cross-section. 

\subsection{Large $x$ Spin Glass Region}

\begin{figure}[tbp]
\includegraphics[width=8cm]{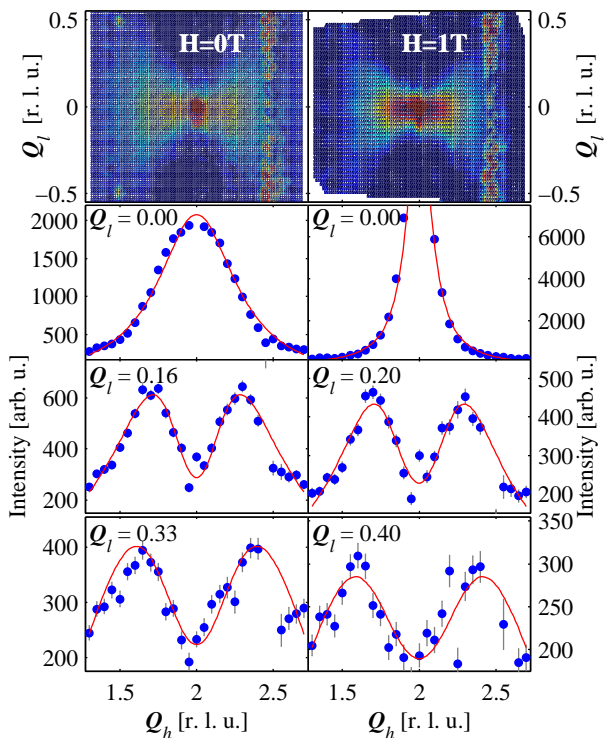}
\caption{(Color online).[Top] $\boldsymbol{Q}$-dependent scattered intensity centered around the (2,0,0) Bragg peak at 50 mK and [left] 0~T, [right] 1~T. The scans show a butterfly shape scattering typical for dipolar coupled Ising moments, with the spin-spin correlation length increasing at 1~T. [Lower panels] $\boldsymbol{Q}_h$ scans across the diffuse scattering for selected $\boldsymbol{Q}_l$ after subtraction of 3~T background to remove copper and aluminum powder lines (these are the origin of the streaks at $\boldsymbol{Q}_h\sim2.6$ in the top panels). The data is subsequently fit to eq's.~\ref{eq:Ising_needle_1} and \ref{eq:Ising_needle_2}.}
\label{50_Rita}
\end{figure}

Measurements on both the $x=0.50$ and $x=0.25$ samples were carried out on the cold neutron triple axis spectrometer RITA-II at PSI in Switzerland. For $x=0.50$ the sample had total dimensions of $12\times 12\times 40$~mm and was mounted in a vertical field superconducting magnet and dilution fridge. The scattering plane was the $(h00)-(00l)$ plane, the magnetic field was aligned along the b-axis. The incoming neutrons along with the analyzer blades were tuned to a wavevector $k_i = 1.97 ~\text{\AA}^{-1}$. The measurements were carried out using all analyzer blades in the so-called \textit{monochromatic imaging mode}, where each analyzer detects a slightly different $\boldsymbol{Q}$-vector~\cite{Lefmann2006RITA} to measure out a region of $\boldsymbol{Q}$-space centered around the (2,0,0) Bragg peak. This region has been mapped out at base temperature, both in zero field and in a 1~T transverse field as shown in Fig~\ref{50_Rita}.

In the top two panels, the $\boldsymbol{Q}$-dependent scattered intensity is seen to have a butterfly shape distribution, as is expected for the critical scattering of dipolar coupled Ising moments~\cite{als-nielsen_1976}. When a transverse field of 1~T is applied, the pattern narrows and increases in intensity near the center, indicating an increase in the spin-spin correlation length. The lower panels show $\boldsymbol{Q}$$_h$ scans taken at selected $\boldsymbol{Q}_l$.

An analytical expression for the magnetic cross-section of Ising moments in the quasi-elastic limit is \cite{Lovesey_1984}:

\begin{equation}
\frac{\mathrm{d}\sigma}{\mathrm{d}\Omega}\propto\left[1-\left(\frac{\boldsymbol{Q}_{z}}{\boldsymbol{Q}}\right)^{2}\right]f^{2}\left(\boldsymbol{Q}\right)\frac{\chi\left(\boldsymbol{Q},T\right)}{\chi^{0}\left(T\right)},\label{eq:Ising_needle_1}
\end{equation}

where $z$ indicates the Ising direction, $f$$\left(\boldsymbol{Q}\right)$ is the magnetic form factor of Ho$^{3+}$, $\chi\left(\boldsymbol{Q},T\right)$ is the wave vector and temperature-dependent susceptibility and $\chi^{0}\left(T\right)$ is the single ion susceptibility of the system. If we consider $\boldsymbol{q}=\boldsymbol{Q}-\tau$, the deviation of $\boldsymbol{Q}$ from the (2,0,0) reciprocal lattice vector, then renormalization-group theory for a uniaxial, dipolar-coupled system gives \citep{aharony_1973}:

\begin{equation}
\frac{1}{\chi\left(\boldsymbol{Q},T\right)}\propto1+\xi^{2}\left[\boldsymbol{q}^{2}+g\left(\frac{\boldsymbol{q}_{z}}{\boldsymbol{q}}\right)^{2}\right],\label{eq:Ising_needle_2}
\end{equation}

where $\xi$ is the in-plane correlation length and $g$ is an anisotropy factor. The $\boldsymbol{Q}_h$ scans shown in Fig.~\ref{50_Rita} have been fit to this form of scattering while allowing both $\xi$ and $g$ to vary. Before the fit was carried out, powder lines from the copper sample holder and aluminum cryostat (these are the origin of the streaks at $\boldsymbol{Q}_h\sim2.6$ in the top panels) were removed from the measured signal. This was done by subtracting a $\boldsymbol{Q}_h$ scan centered around $\boldsymbol{Q}=(2,0,0)$ taken at 3~T from the data. In zero field a correlation length of $\xi=16\pm1\text{\,\AA}$ and an anisotropy factor of $g=3.3\pm0.5\text{\,\AA}^{-2}$ are found. When the transverse field is applied, the in-plane correlation length increases to $\xi=49\pm4\text{\,\AA\ }$ and the anisotropy factor increases to $g=32\pm5\text{\,\AA}^{-2}$. This implies that not only are the clusters growing in size in the field, but their geometry is changing and they are becoming relatively wider.

\subsection{Small $x$ Spin-Glass Region}

\begin{figure}[tbp]
\includegraphics[width=8cm]{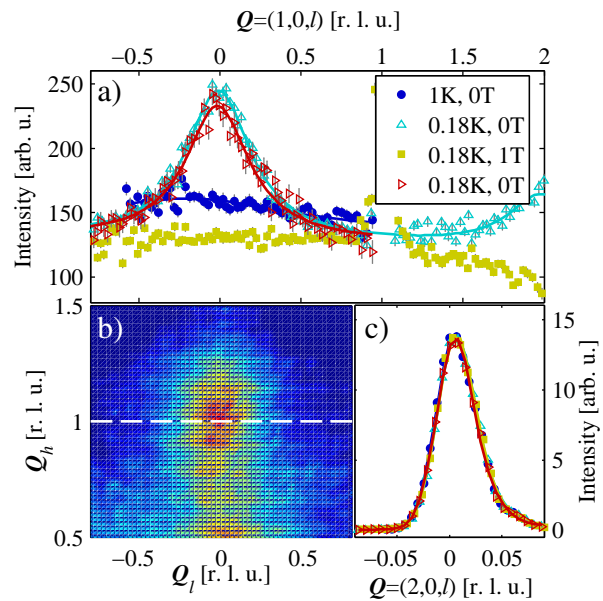}
\caption{(Color Online). Neutron scattering data on LiHo$_{0.25}$Er$_{0.75}$F$_4$. (a) $\boldsymbol{Q}_l$ scans around the (100) forbidden reflection, showing a broad Lorentzian component, corresponding to a spin-spin correlation of $\xi= 44\pm 2~\text{\AA}$ along the c-axis. (b) $\boldsymbol{Q}$ space intensity distribution of this scattering indicating a slight anisotropy with an a-axis correlation length of $\xi= 13\pm 1~\text{\AA}$ along the a-axis. (c) $\boldsymbol{Q}_l$ scans around the (2,0,0) Bragg peak clearly indicating an absence of ferromagnetic correlations.} 
\label{x=0_25_order}
\end{figure}

Measurements on the $x=0.25$ composition were carried out in a dilution fridge and a 1.8~T horizontal magnet along the a-axis and scattering neutrons in the $(h00)-(00l)$ plane. Scans were performed around the nuclear extinct (1,0,0) position along with the (2,0,0) Bragg peak to search for respectively antiferromagnetic and ferromagnetic order and are shown in Fig \ref{x=0_25_order}. The scans were taken at the following fields and temperatures: 1~K and 0~T, 180~mK and 0~T, 180~mK and 1~T, and finally 180~mK and 0~T. 
 
The scans centered around the $(2,0,0)$ Bragg peak position remain both temperature and field independent, indicating that there is no ferromagnetic order in the system. Centered around the $(1,0,0)$ position on the other hand is a very broad and relatively intense Lorentzian signal. If an exponential decay of correlations is assumed then the correlation length is simply the half width half maximum of this Lorentzian:

\begin{equation}
I\propto\frac{1}{1+\boldsymbol{Q}^{2}\xi.^{2}}.
\end{equation}

The fit is made by first subtracting the high field signal and secondly by assuming that the increase in intensity at $\boldsymbol{Q}=(1,0,2)$ is due to the same antiferromagnetic correlations and therefore a Lorentzian with a different amplitude but the same width. Applying this fit yields a c-axis correlation length of $\xi= 44\pm 2~\text{\AA}$. The pseudo-color map in panel (b) shows the $\boldsymbol{Q}$ dependence of this scattering and similar analysis results in an a-axis correlation length of $\xi= 13\pm 1~\text{\AA}$. Taken together, these two pieces of information indicate elliptical regions of correlated moments containing around 30 rare earth sites.

\begin{figure}[tbp]
\includegraphics[width=8cm]{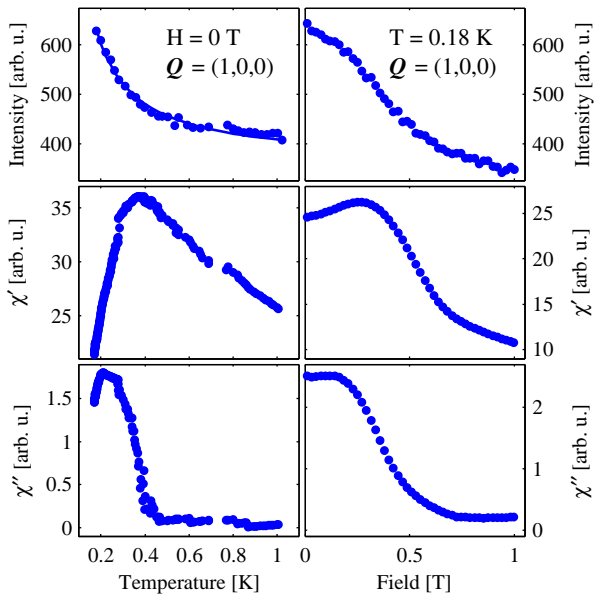}
\caption{(Color online). Zero-field temperature dependence (left) and field dependence at 180~mK (right) of the $\boldsymbol{Q}=(100)$ scattered intensity and simultaneously measured AC susceptibility of LiHo$_{0.25}$Er$_{0.75}$F$_4$.}
\label{x=0_25_T_H_dep}
\end{figure}

The field and temperature dependences of the spin glass were determined by measuring the scattered intensity at $\boldsymbol{Q}=(1,0,0)$ while ramping the field and temperature, respectively. AC susceptibility measurements were carried out \textit{in-situ} simultaneously during the field and temperature scans. For the temperature scans the temperature was ramped with the following ramping rates: $17~\mu$K/s for 50~mK~$ < T < $~500~mK and $42~\mu$~K/s for 500~mK~$\leq T <$~1~K. The temperature scan is presented in the left side of Fig.~\ref{x=0_25_T_H_dep} and the field scan on the right side.

Temperature scans show a continuous decrease in spin-spin correlations which fall off quickly up to $\sim 0.4$~K after which they decrease more slowly (top panel). The cross-over at $0.4$~K coincides with the onset of $\chi^{\prime\prime}$, and the peak in $\chi^{\prime}$, which gives $T_{\text{f}}$, occurs at a lower temperature.  The rather high temperature where spin-spin correlations increase suddenly can be explained by the THz frequency scale of the neutrons and the logarithmic frequency dependence of $T_{\text{f}}$. A quick calculation assuming an Arrhenius dependence up to the THz range and using $T_{\text{f}}$ and $\mathcal{K}$ from Table~\ref{tab_summary} finds a $T_{\text{f}}(1\text{THz})\sim 0.5$~K. In the field scan, the intensity once again drops off continuously. A kink in neutron intensity appears at the same field as the peak in $\chi^{\prime\prime}$.

\section{Discussion\label{Discussion}}

The rich phase diagram of LiHo$_x$Er$_{1-x}$F$_4$ is an ideal playground for the study of spin-glass materials. Like in LiHo$_x$Y$_{1-x}$F$_4$, by substituting holmium, first the Curie temperature of the ferromagnet is decreased. At lower temperatures a spin-glass state emerges, effectively inside of a ferromagnetic matrix. As $x$ is decreased further, at around $x=0.6$ the ferromagnetism is completely suppressed and replaced by a spin-glass state showing broad features in the AC susceptibility. As $x$ decreases towards zero, the features in the AC susceptibility sharpen and have the allure of a canonical-like spin glass. Each of these distinct regions of the phase diagram are discussed in detail in the following paragraphs.

\subsection{Ferromagnetic Region}

Upon first inspection, for large $x$ the system appears to be a good example of a re-entrant spin-glass, where the spin glass state is entered not from a paramagnet but from a ferromagnetic phase. The neutron scattering results on $x=0.79$ indicate that the ferromagnetic state persists down to 50 mK, implying the system is perhaps better described as an \textit{embedded spin-glass}, where the spin-glass co-exists with the ferromagnetic order. 

In general the frequency dependence of this spin-glass like state does not seem to be well described by either Arrhenius or Critical scaling dynamics. In both cases the extracted parameters seem to vary dramatically from one concentration to the next. In the case of critical scaling, the parameters are very far from those expected from a typical spin-glass. For the Arrhenius fits, the low values of $f_0$ appear to indicate that the entities freezing out are clusters rather than individual spins. This seems somewhat unlikely as cluster size seemingly increases as Er content decreases, but could be due to the spins existing with a long-range ordered matrix. 

This embedded spin-glass state shares some similarities with the \textit{anti-glass} state observed in LiHo$_{0.045}$Y$_{0.955}$F$_4$~\cite{Ghosh2002}. More specifically in both compounds there is a clear deviation from Arrhenius law behavior of the spin-glass freezing temperature, but with no evidence of a zero-frequency freezing temperature.

\begin{figure}[tbp]
\includegraphics[width=8.0cm]{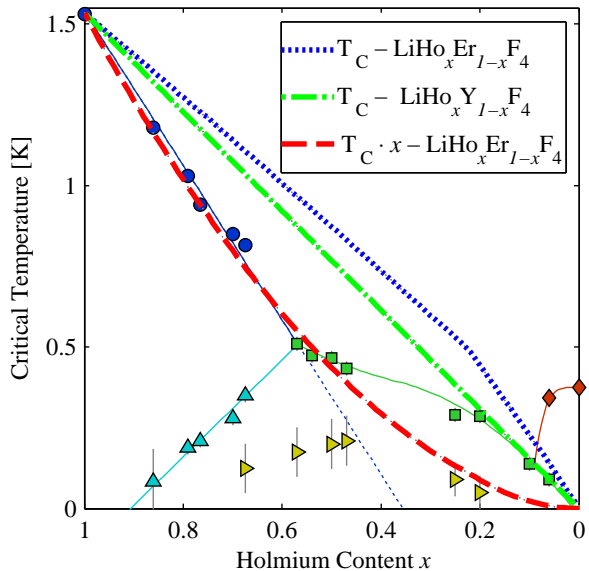}
\caption{(Color Online). Experimental phase diagram of LiHo$_x$Er$_{1-x}$F$_4$ compared with VCMF calculations (blue dotted line). As a reference the VCMF result for LiHo$_x$Y$_{1-x}$F$_4$, which is in good agreement with experiments is shown (green dash-dot line). It appears that multiplying $T_C$ found by VCMF additionally by the Ho content $x$ yields good agreement with the experimental phase diagram (red dashed line).}
\label{phasediag_MF}
\end{figure}

The ferromagnetic Curie temperature $T_C$ decreases linearly and more rapidly than in LiHo$_{x}$Y$_{1-x}$F$_4$. Virtual crystal mean-field (VCMF) calculations fail to capture this effect, giving a $T_C(x)$ \textit{larger} than in LiHo$_{x}$Y$_{1-x}$F$_4$, as can be seen in Fig.~\ref{phasediag_MF}. Given the success of mean-field theory to explain the phase diagram of LiHo$_{x}$Y$_{1-x}$F$_4$, this failing is somewhat surprising. Inhomogeneous mean-field (iMF) calculations on a lattice of 100 x 100 x 100 unit cells gives identical results to the VCMF, implying the effect is not simply due to the disorder in the location of magnetic moments and confirming the system is no longer mean-field. In investigating this reduction of $T_C(x)$, we came across an interesting result. If the $T_C(x)$ obtained from mean-field is multiplied with an additional $x$, then there is a remarkably good agreement with the experimental phase diagram as seen in Fig.~\ref{phasediag_MF}. For the time being we have no theoretical argument that this is more than a coincidence, although it could prove a useful starting point for additional theoretical work.

Further calculations to explain the rapidly decreasing $T_C$ have been carried out. First, the Hamiltonian containing only the two lowest crystal field levels was diagonalized for a small cluster of 8 Ho ions surrounding a central Er ion (4 nearest neighbors, 4 next nearest neighbors) in the presence of the temperature dependent mean-field generated in pure LiHoF$_4$. These calculations gave identical results to the mean-field ones - the Er becomes polarized by the Ho ions, increasing $T_C$. Similar calculations using the full Hamiltonian were carried out for an Er ion with 4 nearest-neighbor Ho ions, giving the same result. As there is a slight change in lattice parameters for LiHoF$_4$ and LiErF$_4$, on the order of 0.5 \%, another possibility is that the distorted lattice could influence the crystal field levels. To investigate this hypothesis, point charge crystal field calculations were carried out for a variety of symmetrical lattice distortions and the resulting crystal field used in VCMF calculations. The net effect of the distortions was a change in $T_C$ far too small to explain the experimental data.

More theoretical work must be carried out in order to understand the dependence of $T_C$ as a function of Er content. It is clear from the calculations already carried out that it is not a local quantum or classical effect and therefore probably a long range effect. Given the success of Classical Monte Carlo simulations~\cite{biltmo_2007,biltmo_2008} in explaining the non-MF behavior observed in LiHo$_x$Y$_{1-x}$F$_4$ for $x<0.5$, these calculations may shed more light on the situation.

\subsection{Large $x$ Spin Glass Region}

As $x$ decreases, the ferromagnetic component in the $\chi^{\prime}$ disappears completely. In $\chi^{\prime\prime}$ on the other hand, the peak due to the spin-glass begins to widen. Given the existence of a spin-glass embedded within the Ferromagnetic state for larger $x$, it seems likely that this widening is due to two physically distinct transitions. Indeed the large difference in the form of $\chi_{xx}$ and $\chi_{zz}$ indicate that at high temperatures an Ising spin-glass forms and at lower temperatures an $XY$ spin-glass forms within this glass. The neutron scattering data on $x=0.50$ confirms the presence of the Ising spin-glass, which consists of elongated needle-like clusters. This kind of spin-glass where at high temperatures the Ising moments freeze out then at lower temperatures the XY moments freeze has been theoretically predicted for mixed anisotropy spin-glasses~\cite{viana_phase_1983}.

\subsection{Small $x$ Spin Glass Region}

Moving to even lower $x$, the susceptibility evolves continuously towards that of a canonical spin glass, where the features in both real and imaginary susceptibility are analogous to those seen in the canonical spin glasses \cite{mydosh1993}. Neutron scattering measurements confirm the presence of short-range correlations as are present in spin glasses. Surprisingly, these correlations are anti-ferromagnetic in nature, not ferromagnetic as could be expected for Ho ions. Mapping the correlations in the scattering plane reveals an ellipsoidal correlation volume with major radius of $44 \pm 2~\text{\AA}$ along the c-axis and minor radius of $13 \pm 1 ~\text{\AA}$ along the a-axis. The field and temperature dependence of the correlations is consistent with previous studies on spin glasses, where there is a monotonic decrease in intensity as a function of temperature and field.

\section{\label{Conclusion}Conclusion}

To conclude, we have measured the experimental phase diagram of LiHo$_x$Er$_{1-x}$F$_4$, finding a rich phase diagram with at least three distinct regions. At large $x$ the system is a ferromagnet with a low-temperature embedded spin glass. For $x\sim0.5$ the system likely consists of two coexisting spin-glasses, one of which freezes at around 0.5 K and the other at around 0.2 K. At lower $x$ the features of the spin-glass sharpen and by $x=0.25$ there is only evidence of a single spin-glass state showing antiferromagnetic spin-spin correlations.

\section*{Acknowledgments}
We gratefully thank J. Jensen and M. Gingras for fruitful discussions. We thank D. Biner, Bern, for the synthesis of the LiHo$_x$Er$_{1-x}$F$_4$ samples. Financial support is gratefully acknowledged from the Swiss SNF, the European Research Council, and DANSCATT of the Danish Agency for Science, Technology, and Innovation. Neutron experiments were performed at the Swiss spallation source (SINQ), Paul Scherrer Institut (PSI), Switzerland and at Helmholtz-Zentrum Berlin (HZB), Germany.

%

\end{document}